# The missing elements in the telegraph equations

**Ben David Golan**[1]
[1]Electrical Engineering Department, University of Ariel, Ariel, Israel

Corresponding author: Ben David Golan (e-mail: ben.golan@msmail.ariel.ac.il).

This work of mine was not supported by anyone, since no one wanted to take it seriously.

**ABSTRACT** The conventional modeling of transmission lines relies on the classical telegraph equations, originally formulated over 150 years ago. These equations are typically derived by representing the line as an assembly of infinitesimal inductive, capacitive, and resistive elements. However, this formulation is fundamentally flawed, as a transmission line cannot be accurately described through a discretized model of infinitesimal lumped components. Instead, a more rigorous approach should derive the governing equations directly from Maxwell's equations in conjunction with Ohm's law. This paper presents such a derivation and introduces a corrected formulation, herein referred to as the Trump Equations.

## I. INTRODUCTION

Today, it is common to analyze what happens on a transmission line using the following equations, known as the "telegraph equations":

$$\frac{\partial}{\partial z}V = -L^*\frac{\partial}{\partial t}I - R^*I$$

$$\frac{\partial}{\partial z}I = -C^*\frac{\partial}{\partial t}V - G^*V$$

By applying straightforward mathematical manipulations to the telegraph equations, one can derive the following second-order differential equation for the voltage:

$$\frac{\partial^2}{\partial z^2}V - L^*C^*\frac{\partial^2}{\partial t^2}V = (R^*C^* + G^*L^*)\frac{\partial}{\partial t}V + G^*R^*V \quad (1)$$

These equations are obtained by dividing the line into infinitesimal sections with resistance, inductance, and capacitance, and analyzing these sections using K.V.L and K.C.L, as shown in the following figure:

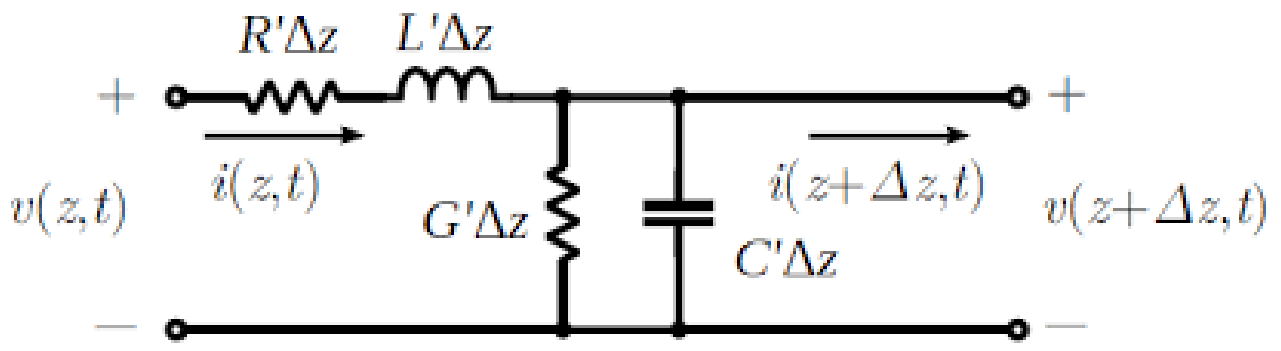

FIGURE 1. **Schematic representation of an infinitesimal segment of a transmission line modeled using distributed resistance, inductance, and capacitance.**

This approach was originally introduced by Oliver Heaviside in his 1876 publication "On the Extra Current"[1] . Since its introduction, the formulation proposed by Heaviside has remained largely unchanged. However, the underlying methodology—dividing the transmission line into infinitesimal lumped components—is, in the present view, fundamentally flawed and does not accurately represent the physical behavior of the line.
In this work, it is demonstrated that a corrected form of the telegraph equations can be derived solely from Maxwell's equations and Ohm's law, without relying on lumped-element assumptions. The derivation employs the following notations, summarized in Table:

| Symbol | Quantity |
|---|---|
| R | Electrical resistance |
| $B$ | magnetic flux density, magnetic induction |
| $E$ | The electric field |
| $V$ | Electrical voltage |
| $I$ | Electric current |
| σ | Electrical conductivity of transmission lines |
| $\sigma_G$ | Electrical conductivity of the dielectric medium between the lines |
| J | Electric current density |
| μ | The permeability constant |
| ε | The dielectric constant |
| Φ | Electric flux |
| Q | Electric charger |

## II. RELATED WORKD

Since Oliver Heaviside first introduced the telegraph equations in his 1876 article "On the Extra Current"[1], the equations have remained essentially unchanged and unchallenged. As a result, this publication stands as the sole foundational work on the subject, with no significant revisions or alternatives proposed in the intervening years. This continued reliance on the original formulation underscores the need to re-examine its assumptions and validity in light of modern electromagnetic theory.

## III. development of the first telegraph equation

A rectangular frame between transmission lines, as illustrated in the following figure:

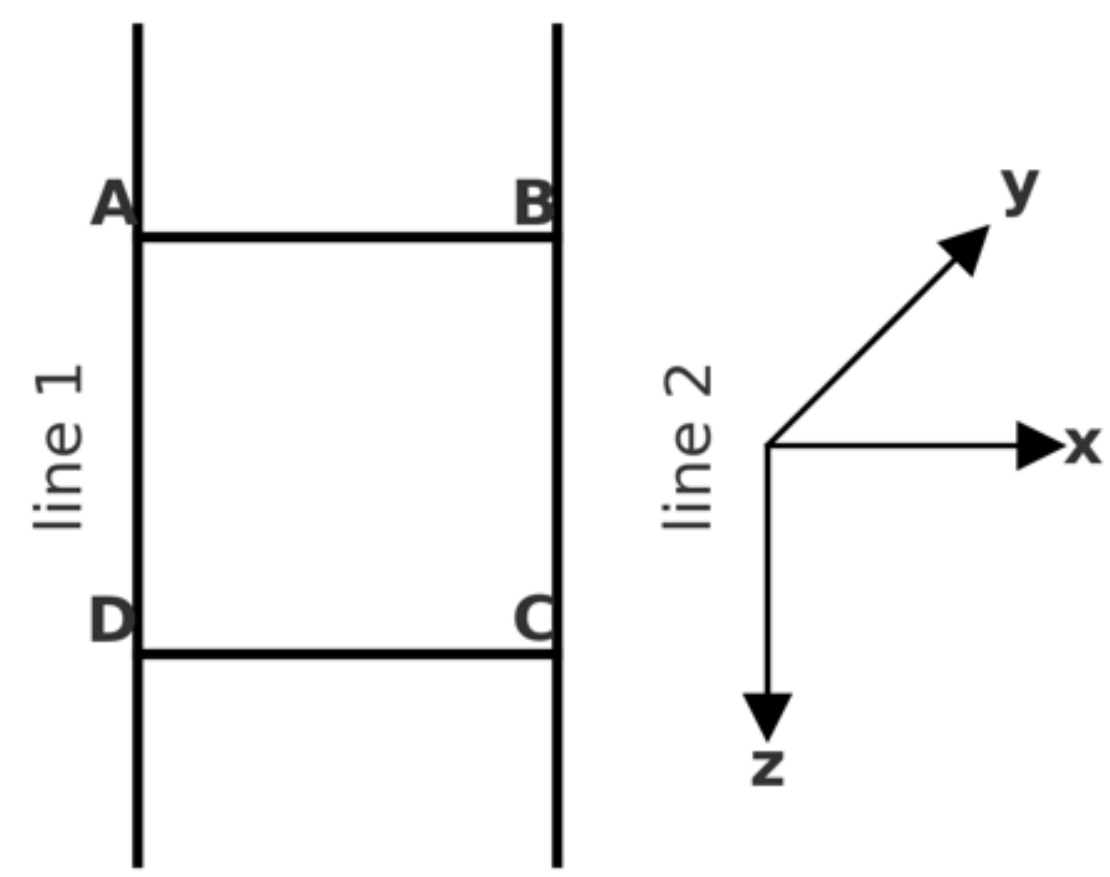

FIGURE 2. **Transmission lines with imaginary frame ABCD**

According to Faraday's law:

$$\nabla \times E = -\frac{\partial B}{\partial t}$$

We run CURL on both sides and get:

$$\nabla \times \nabla \times E = -\frac{\partial}{\partial t}\nabla \times \mathrm{B} \qquad (2)$$

According to Ampere's law:

$$\mu\mathrm{J} + \mu\varepsilon\frac{\partial E}{\partial t} = \nabla \times \mathrm{B}$$

Differentiating both sides of the equation with respect to time yields:

$$\frac{\partial}{\partial t}\left(\mu\mathrm{J} + \mu\varepsilon\frac{\partial E}{\partial t}\right) = \frac{\partial}{\partial t}\nabla \times \mathrm{B} \qquad (3)$$

Equations (2) and (3) are combined to yield the following result:

$$\frac{\partial}{\partial t}(\mu J + \mu\varepsilon\frac{\partial E}{\partial t}) = -\nabla\times\nabla\times E$$

A line integral of the second kind is performed on both sides of the equation along the closed contour ABCD:

$$\oint_{ABCD}\frac{\partial}{\partial t}(\mu J + \mu\varepsilon\frac{\partial E}{\partial t})dl = \oint_{ABCD} -\nabla\times\nabla\times E\, dl$$

The time derivative is taken outside the integral, assuming the integrand is sufficiently smooth:

$$\frac{\partial}{\partial t}\oint_{ABCD}(\mu J + \mu\varepsilon\frac{\partial E}{\partial t})dl = \oint_{ABCD} -\nabla\times\nabla\times E\, dl$$

The line integral on the left-hand side is decomposed into its individual segments:

$$\frac{\partial}{\partial t}[\oint_A^B(\mu J + \mu\varepsilon\frac{\partial E}{\partial t})dl + \oint_C^D(\mu J + \mu\varepsilon\frac{\partial E}{\partial t})dl + \oint_B^C(\mu J + \mu\varepsilon\frac{\partial E}{\partial t})dl + \oint_D^A(\mu J + \mu\varepsilon\frac{\partial E}{\partial t})dl] = \oint_{ABCD} -\nabla\times\nabla\times E\, dl \quad (4)$$

According to Ohm's law: $E = \frac{J}{\sigma}$ , Substituting this relation into equation (4) yields:

$$\frac{\partial}{\partial t}[\oint_A^B(\mu\sigma E + \mu\varepsilon\frac{\partial E}{\partial t})dl + \oint_C^D(\mu\sigma E + \mu\varepsilon\frac{\partial E}{\partial t})dl + \oint_B^C(\mu J + \frac{\mu\varepsilon}{\sigma}\frac{\partial J}{\partial t})dl + \oint_D^A(\mu J + \frac{\mu\varepsilon}{\sigma}\frac{\partial J}{\partial t})dl] = \oint_{ABCD} -\nabla\times\nabla\times E\, dl \quad (5)$$

where σ denotes the electrical conductivity of the transmission line.

but: $V_{AB} = \int_A^B E\, dl$ , $-V_{DC} = V_{CD} = \int_C^D E\, dl$

Substituting this expression into equation (5) yields:

$$\frac{\partial}{\partial t}[\mu\sigma V_{AB} + \mu\varepsilon\frac{\partial V_{AB}}{\partial t} - \mu\sigma V_{DC} - \mu\varepsilon\frac{\partial V_{DC}}{\partial t} + \oint_B^C(\mu J + \frac{\mu\varepsilon}{\sigma}\frac{\partial J}{\partial t})dl + \oint_D^A(\mu J + \frac{\mu\varepsilon}{\sigma}\frac{\partial J}{\partial t})dl] = \oint_{ABCD} -\nabla\times\nabla\times E\, dl$$

$$\to \frac{\partial}{\partial t}[\mu\sigma(V_{AB} - V_{DC}) + \mu\varepsilon\frac{\partial(V_{AB} - V_{DC})}{\partial t} + \oint_B^C(\mu J + \frac{\mu\varepsilon}{\sigma}\frac{\partial J}{\partial t})dl + \oint_D^A(\mu J + \frac{\mu\varepsilon}{\sigma}\frac{\partial J}{\partial t})dl] = \oint_{ABCD} -\nabla\times\nabla\times E\, dl \quad (6)$$

Along segments DA and BC the current density is given by: $J = \frac{I}{S}$, also: $\int_D^A \quad = -\int_A^D$

Substituting this expression into equation (6) yields:

$$\frac{\partial}{\partial t}[\mu\sigma(V_{AB} - V_{DC}) + \mu\varepsilon\frac{\partial(V_{AB} - V_{DC})}{\partial t} + \oint_B^C(\frac{\mu}{S}I_2 + \frac{\mu\varepsilon}{\sigma S}\frac{\partial I_2}{\partial t})dl - \oint_A^D(\frac{\mu}{S}I_1 + \frac{\mu\varepsilon}{\sigma S}\frac{\partial I_1}{\partial t})dl] = \oint_{ABCD} -\nabla\times\nabla\times E\, dl$$

Taking the limit as the frame width Δz→0 we obtain:

$$\frac{\partial}{\partial t}[\mu\sigma(V_{AB} - V_{DC}) + \mu\varepsilon\frac{\partial(V_{AB} - V_{DC})}{\partial t} + (\frac{\mu}{S}I_2 + \frac{\mu\varepsilon}{\sigma S}\frac{\partial I_2}{\partial t})\Delta z - (\frac{\mu}{S}I_1 + \frac{\mu\varepsilon}{\sigma S}\frac{\partial I_1}{\partial t})\Delta z] = \oint_{ABCD} -\nabla\times\nabla\times E\, dl$$

Upon division by Δz, the equation becomes:

$$\frac{\partial}{\partial t}[\mu\sigma\frac{V_{AB} - V_{DC}}{\Delta z} + \mu\varepsilon\frac{\partial}{\partial t}\frac{(V_{AB} - V_{DC})}{\Delta z} + \frac{\mu}{S}(I_2 - I_1) + \frac{\mu\varepsilon}{\sigma S}(\frac{\partial(I_2 - I_1)}{\partial t})] = \frac{1}{\Delta z}\oint_{ABCD} -\nabla\times\nabla\times E\, dl$$

$$\to \frac{\partial}{\partial t}[\mu\sigma\frac{\partial V}{\partial z} + \mu\varepsilon\frac{\partial}{\partial t}\frac{\partial V}{\partial z} + \frac{\mu}{S}(I_2 - I_1) + \frac{\mu\varepsilon}{\sigma S}(\frac{\partial(I_2 - I_1)}{\partial t})] = \frac{1}{\Delta z}\oint_{ABCD} -\nabla\times\nabla\times E\, dl$$

It is important to note that: $I_2 - I_1 \neq 0$ , as the currents in the two conductors of the transmission line flow in opposite directions. Consequently, the corresponding current densities satisfy: $J_2 = -J_1$ , and the magnitude of the current in each conductor is given by: $I = JS$. Assuming that the transmission line is uniform along its length, the magnitudes of the currents at corresponding cross-sections are equal, i.e: $I_2 - I_1 = 2I_1 = -2I_2 = 2I$ , Hence:

$$\to \frac{\partial}{\partial t}\left[\mu\sigma\frac{\partial V}{\partial z} + \mu\varepsilon\frac{\partial}{\partial t}\frac{\partial V}{\partial z} + \frac{2\mu}{S}I + \frac{2\mu\varepsilon}{\sigma S}\frac{\partial I}{\partial t}\right] = \frac{1}{\Delta z}\oint_{ABCD} -\nabla\times\nabla\times E\, dl \quad (7)$$

We now turn to the right-hand side of the equation. According to Stokes' theorem:

$$\frac{1}{\Delta z}\oint_{ABCD} -\nabla\times\nabla\times E\, dl = -\frac{1}{\Delta z}\iint_{ABCD}\nabla\times\nabla\times\nabla\times E\ ds$$

Since ABCD is a rectangular loop lying in the XZ-plane and oriented perpendicular to the Y-axis, the surface integral can be expressed as the following double integral:

$$\frac{1}{\Delta z}\oint_{ABCD} -\nabla\times\nabla\times E\, dl = -\frac{1}{\Delta z}\iint_{ABCD}\nabla\times\nabla\times\nabla\times E\ ds = -\frac{1}{\Delta z}\iint_{ABCD}(\nabla\times\nabla\times\nabla\times E)_y\ dxdz$$

Due to the small width of the frame, the surface can be approximated as a sum of differential rectangular elements, as shown in the accompanying figure:

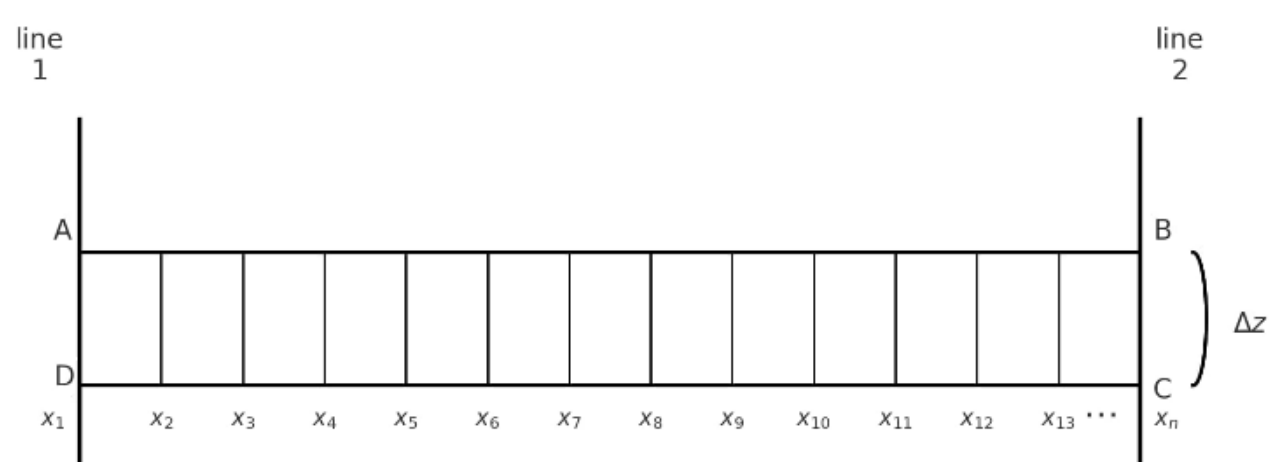

FIGURE 3. **Transmission lines with imaginary frame ABCD**

Therefore:

$$\frac{1}{\Delta z}\oint_{ABCD} -\nabla\times\nabla\times E\, dl = -\frac{1}{\Delta z}\iint_{ABCD}\nabla\times\nabla\times\nabla\times E\ ds = -\frac{1}{\Delta z}\iint_{ABCD}(\nabla\times\nabla\times\nabla\times E)_y\, dxdz = -\frac{1}{\Delta z}\sum_1^n \Delta x_n\Delta z\,(\nabla\times\nabla\times\nabla\times E)_y = -\sum_1^n \Delta x_n\frac{\Delta z}{\Delta z}(\nabla\times\nabla\times\nabla\times E)_y =$$

$$-\sum_{1}^{n} \Delta x_n \, (\nabla \times \nabla \times \nabla \times E)_y = -\int_A^B (\nabla \times \nabla \times \nabla \times E)_y \, dx$$

Therefore, the resulting expression is given by:

$$\frac{1}{\Delta z} \oint_{ABCD} -\nabla \times \nabla \times E \, dl = -\int_A^B (\nabla \times \nabla \times \nabla \times E)_y \, dx$$

As is known:
$\nabla \times \nabla \times E = \nabla\nabla \cdot E - \nabla^2 E$

it follows that:
$\frac{1}{\Delta z}\oint_{ABCD} -\nabla \times \nabla \times E \, dl = -\int_A^B (\nabla \times \nabla \times \nabla \times E)_y \, dx =$
$-\int_A^B (\nabla \times \nabla\nabla \cdot E - \nabla \times \nabla^2 E)_y \, dx =$

However, since the curl of a gradient is identically zero, it follows that: $\nabla \times \nabla\nabla \cdot E = 0$

$\rightarrow \frac{1}{\Delta z}\oint_{ABCD} -\nabla \times \nabla \times E \, dl = -\int_A^B (\nabla \times \nabla \times \nabla \times E)_y \, dx =$
$\int_A^B (\nabla \times \nabla^2 E)_y \, dx$

The Y-component of the expression inside the parentheses is written explicitly as follows:

$\frac{1}{\Delta z}\oint_{ABCD} -\nabla \times \nabla \times E \, dl = -\int_A^B (\nabla \times \nabla \times \nabla \times E)_y \, dx =$
$\int_A^B (\nabla \times \nabla^2 E)_y \, dx = \int_A^B \frac{\partial}{\partial z}(\nabla^2 E_x) - \frac{\partial}{\partial x}(\nabla^2 E_z) \, dx$

Consequently, the resulting expression is:

$$\frac{1}{\Delta z} \oint_{ABCD} -\nabla \times \nabla \times E \, dl = \int_A^B \frac{\partial}{\partial z}(\nabla^2 E_x) - \frac{\partial}{\partial x}(\nabla^2 E_z) \, dx$$

However, since the wave propagating along a transmission line is of the Transverse Electromagnetic (TEM) type, it follows that : $E_z = 0$

$$\rightarrow \frac{1}{\Delta z}\oint_{ABCD} -\nabla \times \nabla \times E \, dl = \int_A^B \frac{\partial}{\partial z}(\nabla^2 E_x) \, dx = \int_A^B \frac{\partial}{\partial z}\left(\frac{\partial^2}{\partial x^2}E_x + \frac{\partial^2}{\partial y^2}E_x + \frac{\partial^2}{\partial z^2}E_x\right) \, dx \quad (8)$$

We now demonstrate that the second spatial derivatives of $E_x$ with respect to both x and y vanish, namely: $\frac{\partial^2}{\partial x^2}E_x = \frac{\partial^2}{\partial y^2}E_x = 0$

According to Faraday's law: $\nabla \times E = -\frac{\partial B}{\partial t}$

We focus next on the Z-component of the equation under consideration:

$$(\nabla \times E)_z = -\frac{\partial B_z}{\partial t} \rightarrow \frac{\partial}{\partial x}E_y - \frac{\partial}{\partial y}E_x = -\frac{\partial B_z}{\partial t} \quad (9)$$

Given that the transmission line supports TEM-mode propagation, the associated field configuration is depicted in the following figure:

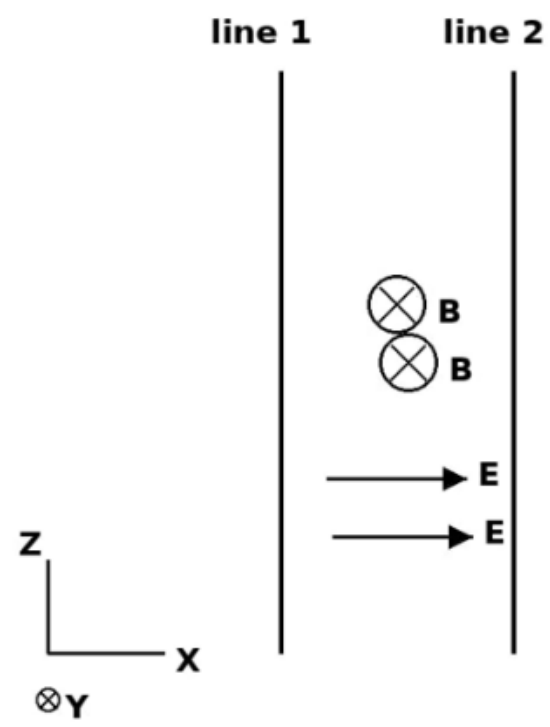

FIGURE 4 . **Field Direction Diagram**

Therefore, since none of the fields possesses a component in the Z-direction, it follows that: $B_z = 0$ , Moreover, as the electric field lacks a component in the Y -direction, we have: $E_y = 0$. Substituting these results into equation (9) yields:

$$\frac{\partial}{\partial y}E_x = 0 \rightarrow \frac{\partial^2}{\partial y^2}E_x = 0 \quad (10)$$

Furthermore Substitution of $B_z = 0$ and (10) into (9) yields:
$\frac{\partial}{\partial x}E_y = 0 \quad (11)$

This completes the first part of the proof.

We now proceed to demonstrate that: $\frac{\partial^2}{\partial x^2}E_x = 0$:
In accordance with Gauss's Law for electric fields:

$$\frac{\partial}{\partial x}E_x + \frac{\partial}{\partial y}E_y + \frac{\partial}{\partial z}E_z = \frac{\rho}{\varepsilon}$$

However, since the wave propagating along a transmission line is of the Transverse Electromagnetic (TEM) type, it follows that : $E_z = 0$

$$\rightarrow \frac{\partial}{\partial x}E_x + \frac{\partial}{\partial y}E_y = \frac{\rho}{\varepsilon}$$

Differentiating the equation with respect to x yields:

$$\rightarrow \frac{\partial^2}{\partial x^2}E_x + \frac{\partial}{\partial x}\frac{\partial}{\partial y}E_y = \frac{\partial}{\partial x}\frac{\rho}{\varepsilon}$$

Reversing the order of differentiation yields:

$$\rightarrow \frac{\partial^2}{\partial x^2}E_x + \frac{\partial}{\partial y}\left(\frac{\partial}{\partial x}E_y\right) = \frac{\partial}{\partial x}\frac{\rho}{\varepsilon}$$

Since, according to (11), $\frac{\partial}{\partial x}E_y = 0$ , it follows that:
$\frac{\partial^2}{\partial x^2}E_x = \frac{\partial}{\partial x}\frac{\rho}{\varepsilon}$

Consequently, since the dielectric material between the conductors contains virtually no net charge, we may assume ρ =0, yielding:

$$\frac{\partial^2}{\partial x^2}E_x = 0 \quad (12)$$

This completes the second part of the proof. We now substitute equations (10) and (12) into equation (8), yielding: $\frac{1}{\Delta z}\oint_{ABCD} -\nabla\times\nabla\times E\,dl = \int_A^B \frac{\partial}{\partial z}(\nabla^2 E_x)\,dx = \int_A^B \frac{\partial}{\partial z}\left(0+0+\frac{\partial^2}{\partial z^2}E_x\right)\,dx = \int_A^B \left(\frac{\partial^3}{\partial z^3}E_x\right)\,dx$

Accordingly, the resulting equation is:

$$\frac{1}{\Delta z}\oint_{ABCD} -\nabla\times\nabla\times E\,dl = \int_A^B \left(\frac{\partial^3}{\partial z^3}E_x\right)\,dx = \frac{\partial^3}{\partial z^3}\int_A^B E_x\,dx = \frac{\partial^3}{\partial z^3}V \qquad (13)$$

By inserting equation (13) into equation (7), the following expression is derived:

$$\rightarrow \frac{\partial}{\partial t}\left[\mu\sigma\frac{\partial V}{\partial z} + \mu\varepsilon\frac{\partial}{\partial t}\frac{\partial V}{\partial z} + \frac{2\mu}{S}I + \frac{2\mu\varepsilon}{\sigma S}\frac{\partial I}{\partial t}\right] = \frac{\partial^3}{\partial z^3}V$$

Transposing terms and multiplying through by $\frac{1}{\mu\sigma}$ the equation becomes:

$$\frac{1}{\mu\sigma}\frac{\partial^3}{\partial z^3}V - \frac{\partial}{\partial t}\left[\frac{\partial V}{\partial z} + \frac{\varepsilon}{\sigma}\frac{\partial}{\partial t}\frac{\partial V}{\partial z} + \frac{2}{\sigma S}I + \frac{2\varepsilon}{\sigma^2 S}\frac{\partial I}{\partial t}\right] = 0$$

Since: $R = \frac{L}{\sigma S} \rightarrow \frac{R}{L} = \frac{1}{\sigma S}$ , it follows that: $\frac{1}{\sigma S}$ represents the resistance per unit length. We therefore define: $\frac{1}{\sigma S} = R^*$

$$\frac{1}{\mu\sigma}\frac{\partial^3}{\partial z^3}V - \frac{\partial}{\partial t}\left[\frac{\partial V}{\partial z} + \frac{\varepsilon}{\sigma}\frac{\partial}{\partial t}\frac{\partial V}{\partial z} + 2R^*I + \frac{2\varepsilon R^*}{\sigma}\frac{\partial I}{\partial t}\right] = 0$$

This equation will be referred to as the First Trump Equation, in honor of the President of the United States, Donald J. Trump.

The resulting equation clearly differs from the standard telegraph equation currently in widespread use:

$$\frac{\partial}{\partial z}V = -L^*\frac{\partial}{\partial t}I - R^*I$$

## IV. Developing the Second Trump Equation

Let's look at the following figure**:**

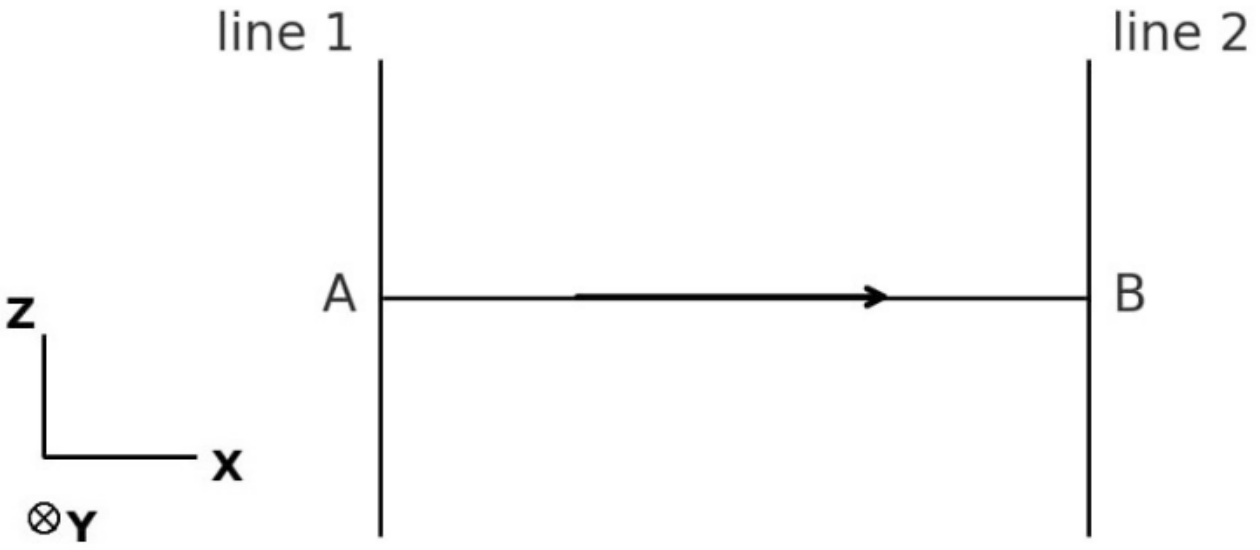

FIGURE 5 . **Transmission lines**

According to Ampère's law, the magnetic field satisfies: $\mu J + \mu\varepsilon\frac{\partial E}{\partial t} = \nabla\times B$

By applying Ohm's law, $J = \sigma E$ we express the current density in terms of the electric field, and Ampère's law becomes:

$$\mu\sigma_g E + \mu\varepsilon\frac{\partial E}{\partial t} = \nabla\times B$$

Here, $\sigma_g$ denotes the electrical conductivity of the dielectric medium.

We now perform a line integral of the second kind on both sides of the equation along the path from point A to point B:

$$\int_A^B \left(\mu\sigma_g E + \mu\varepsilon\frac{\partial E}{\partial t}\right)dl = \int_A^B \nabla\times B\,dl$$

$$\rightarrow \mu\sigma_g\int_A^B E\,dl + \mu\varepsilon\frac{\partial}{\partial t}\int_A^B E\,dl = \int_A^B \nabla\times B\,dl$$

$$\rightarrow \mu\sigma_g V + \mu\varepsilon\frac{\partial V}{\partial t} = \int_A^B \nabla\times B\,dl$$

Given that the line integral is taken exclusively along the X-axis, only the X-component of the vector field contributes, and thus:

$$\rightarrow \mu\sigma_g V + \mu\varepsilon\frac{\partial V}{\partial t} = \int_A^B (\nabla\times B)_x\,dl$$

The X-component of the curl operator is written explicitly as:

$$\rightarrow \mu\sigma_g V + \mu\varepsilon\frac{\partial V}{\partial t} = \int_A^B \frac{\partial B_z}{\partial y} - \frac{\partial B_y}{\partial z}\,dl$$

Given that the wave propagates in the TEM mode, the magnetic field is entirely transverse, implying: $B_z = 0$ , Hence:

$$\rightarrow \mu\sigma_g V + \mu\varepsilon\frac{\partial V}{\partial t} = \int_A^B -\frac{\partial B_y}{\partial z}\,dl \quad (14)$$

According to the differential form of Faraday's law:

$\nabla \times E = -\frac{\partial B}{\partial t}$

We now consider the Y-component of the equation for further analysis:

$(\nabla \times E)_y = -\frac{\partial B_y}{\partial t}$

$\rightarrow \frac{\partial E_x}{\partial z} - \frac{\partial E_z}{\partial x} = -\frac{\partial B_y}{\partial t}$

However, since the wave propagates in the TEM mode, the magnetic field has no component in the Z-direction. Therefore: $E_z = 0$

$\rightarrow \frac{\partial E_x}{\partial z} = -\frac{\partial B_y}{\partial t} \;/\; \int dt$

$\rightarrow -B_y = \int \frac{\partial E_x}{\partial z} dt + f(x, y, z)$ (15)

It is important to note that, since the integration was performed with respect to time, the resulting function f depends only on the spatial coordinates X,Y,Z.

Inserting equation (15) into equation (14), we obtain the following expression:

$\rightarrow \mu\sigma_g V + \mu\varepsilon\frac{\partial V}{\partial t} = \int_A^B \frac{\partial}{\partial z}[\int \frac{\partial E_x}{\partial z} dt + f(x, y, z)]\, dl$

In order to remove the time integration, we take the derivative of both sides of the equation with respect to t:

$\rightarrow \frac{\partial}{\partial t}(\mu\sigma_g V + \mu\varepsilon\frac{\partial V}{\partial t}) = \int_A^B \frac{\partial}{\partial z}[\frac{\partial}{\partial t}\int \frac{\partial E_x}{\partial z} dt + \frac{\partial}{\partial t} f(x, y, z)]dl$

As f does not depend on time, we have $\frac{\partial}{\partial t} f(x, y, z) = 0$, and thus:

$\rightarrow \frac{\partial}{\partial t}(\mu\sigma_g V + \mu\varepsilon\frac{\partial V}{\partial t}) = \int_A^B \frac{\partial}{\partial z}\frac{\partial E_x}{\partial z}\, dl$

$\rightarrow \frac{\partial}{\partial t}(\mu\sigma_g V + \mu\varepsilon\frac{\partial V}{\partial t}) = \frac{\partial^2}{\partial z^2}\int_A^B E_x\, dl$

$\rightarrow \frac{\partial}{\partial t}(\mu\sigma_g V + \mu\varepsilon\frac{\partial V}{\partial t}) = \frac{\partial^2 V}{\partial z^2}$

This equation will be referred to as the Second Trump Equation, in honor of the President of the United States, Donald J. Trump.

It is clearly distinguishable from equation (1), which describes the voltage behavior as derived from the classical telegraph equations:

$$\frac{\partial^2}{\partial z^2}V - L^*C^*\frac{\partial^2}{\partial t^2}V = (R^*C^* + G^*L^*)\frac{\partial}{\partial t}V + G^*R^*V$$

Observe that in the limiting case of a perfect insulator, characterized by (G*=0, R*=0 , $\sigma_g = 0$) the derived equations simplify to:

$$\mu\varepsilon\frac{\partial^2 V}{\partial t^2} = \frac{\partial^2 V}{\partial z^2} \quad (16)$$

The differentiation of the equation with respect to Z yields:

$$\mu\varepsilon\frac{\partial^3 V}{\partial t^2 \partial z} = \frac{\partial^3 V}{\partial z^3}$$

Although substituting this result into the first Trump equation yields:

$$\frac{\varepsilon}{\sigma}\frac{\partial^3 V}{\partial t^2 \partial z} - \frac{\partial}{\partial t}[\frac{\partial V}{\partial z} + \frac{\varepsilon}{\sigma}\frac{\partial}{\partial t}\frac{\partial V}{\partial z} + 2R^*I + \frac{2\varepsilon R^*}{\sigma}\frac{\partial I}{\partial t}] = 0$$

$$\rightarrow \frac{\varepsilon}{\sigma}\frac{\partial^3 V}{\partial t^2 \partial z} - \frac{\varepsilon}{\sigma}\frac{\partial^3 V}{\partial t^2 \partial z} - \frac{\partial}{\partial t}\left[\frac{\partial V}{\partial z} + 2R^*I + \frac{2\varepsilon R^*}{\sigma}\frac{\partial I}{\partial t}\right] = 0$$

$$\rightarrow -\frac{\partial}{\partial t}\left[\frac{\partial V}{\partial z} + 2R^*I + \frac{2\varepsilon R^*}{\sigma}\frac{\partial I}{\partial t}\right] = 0 \;/ \int dt$$

$$\rightarrow -\left[\frac{\partial V}{\partial z} + 2R^*I + \frac{2\varepsilon R^*}{\sigma}\frac{\partial I}{\partial t}\right] = g\,(x, y, z)$$

Given the assumption that $g\,(x, y, z) = 0$ the resulting expression is:

$$\rightarrow \frac{\partial V}{\partial z} = -2R^*I - \frac{2\varepsilon R^*}{\sigma}\frac{\partial I}{\partial t}$$

This is indeed the first telegraph equation. however, it is important to recall that this equation holds under the approximation of $\sigma_g = 0$ ,which was used to derive equation (16).

The two equations developed in this work will henceforth be referred to collectively as the Trump Equations.

Therefore, in order to analyze transmission lines using the Trump equations, one must first solve the Second Trump Equation for the voltage, subject to the appropriate boundary conditions.

The resulting voltage solution is then substituted into the First Trump Equation, yielding the corresponding current distribution along the line.

## V. CONCLUSION

The analysis presented above leads to the conclusion that the classical telegraph equations do not provide a complete description of the physical behavior of transmission lines.

Discrepancies between experimental results and theoretical predictions are often attributed to noise or secondary effects. However, the primary source of these deviations may lie in the fundamental incompleteness of the conventional models themselves.

Consequently, accurate analysis of transmission lines should be based on the Trump Equations derived herein. This insight suggests a need to revisit and revise the theoretical framework taught in academic courses and reflected in standard textbooks on this subject.

## REFERENCES

[1] B.R. Gossick. (1976) Heaviside and Kelvin: A study in contrasts. Annals of Science 33:3, pages 275-287.

**Ben David Golan** is an electrical engineering student at Ariel University. He independently conducted a year-and-a-half-long investigation into the classical telegraph equations, driven by concerns regarding the rigor and completeness of their commonly accepted derivation.

Despite initial skepticism from experienced academics, he remained committed to the development of an alternative theoretical framework. His persistence and belief in the validity of his model ultimately led to the formulation of the Trump Equations, proposed here as a more comprehensive representation of transmission line behavior.